\documentclass[pre,aps,superscriptaddress,showpacs]{revtex4}

\def \beq{\begin{equation}}         \def \eeq{\end{equation}}
\def \beqa{\begin{eqnarray}}        \def \eeqa{\end{eqnarray}}
\def \bea{\begin{array}}        \def \eea{\end{array}}

\usepackage{amsmath}
\usepackage{epsfig}
\usepackage[dvips]{color}

\begin{document}

\title{Derivation of quantum work equalities using quantum Feynman-Kac formula}
\author{Fei Liu}
\email[Email address: ]{feiliu@buaa.edu.cn} \affiliation{School
of Physics and Nuclear Energy Engineering, Beihang University,
Beijing 100191, China}
\date{\today}

\begin{abstract}
{On the basis of a quantum mechanical analogue of the famous
Feynman-Kac formula and the Kolmogorov picture, we present a
novel method to derive nonequilibrium work equalities for
isolated quantum systems, which include the Jarzynski equality
and Bochkov-Kuzovlev equality. Compared with previous methods
in the literature, our method shows higher similarity in form
to that deriving the classical fluctuation relations, which
would give important insight when exploring new quantum
fluctuation relations. }
\end{abstract}
 \maketitle

Feynman-Kac (FK) formula originally found by Feynman in quantum
mechanics~\cite{Feynman} and extended by Kac~\cite{Kac}
establishes an important connection between partial differential
equations  and classical stochastic processes. Briefly, assuming
that in a continuous diffusion process the probability of a
stochastic trajectory $X$ started from a state $x'$ at time $t'$
is $P[X|x',t']$. The solution $u(x',t')$ of the following partial
differential equation
\begin{eqnarray}
\left\{%
\begin{array}{ll}
\partial_{t'} u(x',t')=-{\cal L}^+(x',t')u(x',t')-g(x',t')u(x',t'),\\
u(x',t'=t)=q(x')
   \end{array}
\right.
\end{eqnarray}
has a concise path integral representation~\cite{Stroock}:
\begin{eqnarray}~\label{FKformula}
u(x',t')=\int {\cal D}X e^{ \int_{t'}^tg(x_\tau,\tau)d\tau
}q(x_t,t) P[X|x',t'],
\end{eqnarray}
where ${\cal L}^+$ is the Markovian generator of the diffusion.
This is the famous FK formula in classical stochastic
processes.

Pioneered by Lebowitz and Sphon~\cite{Lebowitz}, this formula
was also found very useful in studying fluctuation
relations~\cite{Hummer01,Chetrite,GeJiang,LiuFPRE09,
LiuFJPA10}. In the past two decades, these important relations
have greatly deepened our understanding about the second law of
thermodynamics and nonequilibrium physics of small
systems~\cite{Evans,Gallavotti,Kurchan,Lebowitz,Bochkov77,JarzynskiPRL97,
Crooks99,HatanoSasa,Maes,SeifertPRL05,Speck,Bustamante}.
Recently, finding quantum fluctuation relations is attracting
intensive interests. Fruitful theoretical and experimental
results~\cite{Kurchanquantum,Yukawa,
Tasaki,Esposito,Campisi,TalknerPRE07,TalknerJPA07,AndrieuxPrl08,CampisiPTRS11,Nakamura}
have been obtained. To our best knowledge, however, there is no
work explicitly using the FK formula. At first glance, the
reason is very obvious, because the classical trajectory
picture on which the FK is based is not available in quantum
physics. Contrary to the intuition, in this Rapid Communication
we use an isolated quantum system as an example to show that
there indeed exists a quantum mechanical analogue of the
classical FK formula and it is very useful to derive the
quantum nonequilibrium work relations including the
Jarzynski~\cite{Tasaki,Kurchanquantum,TalknerPRE07} and
Bochkov-Kuzovlev equalities~\cite{CampisiPTRS11}. \\

{\it Kolmogorov picture and backward invariable.} We start by
introducing essential notations and a new picture that is a
quantum-mechanical analogue of Kolmogorov's
idea~\cite{Kolmogorov} in classical stochastic theory. Although
the picture is virtually equivalent to other well-known
pictures, e.g. the Heisenberg picture, we will see later that
it is very relevant with the time reversal concept. We assume
that the closed quantum system is described by a time-dependent
Hamiltonian $H(t)$. The system's density operators $\rho$ at
two different times $t$ and $t'$ ($<t$) are connected by the
time-evolution operator $U(t')$, i.e.,
\begin{eqnarray}
\label{DensOpera2times}
\rho(t)= U(t)U^{\dag}(t')\rho(t')U(t')U^{\dag}(t).
\end{eqnarray}
Given an arbitrary observable $F$ that does not depend
explicitly on time, we define its Kolmogorov picture as
\begin{eqnarray}
F(t,t')=U(t')F^{\rm H}(t)U^{\dag}(t'),\label{KsPictObs}
\end{eqnarray}
where the superscript H denotes the Heisenberg picture: $F^{\rm
H}(t)$=$U^{\dag}(t)F U(t)$. On the basis of
Eqs.~(\ref{DensOpera2times}) and~(\ref{KsPictObs}), the
expectation value $\langle F\rangle(t)$ at time $t$ in the
picture is
\begin{eqnarray}
{\rm Tr}[F\rho(t)]={\rm
Tr}[F(t,t')\rho(t')],\label{expectationvalule}
\end{eqnarray}
and the equation of motion for $F(t,t')$ with respect to $t'$
is simply
\begin{eqnarray}
\label{KsOperaMotionEq}
\left\{%
\begin{array}{ll}
i\hbar\partial_{t'}F(t,t')=-[F(t,t'), H(t')],\\
F(t,t'=t)=F.
   \end{array}
\right.
\end{eqnarray}
We see that it is a terminal condition rather than initial
condition problem. It is worth pointing out that
Eq.~(\ref{KsOperaMotionEq}) is very distinct from the motion
equation of the same $F(t,t')$ with respect to the forward time
$t$ if the Hamiltonian explicitly depends on time.

Equation~(\ref{expectationvalule}) has a trivial property: the
time derivatives on both sides with respect to $t'$ vanishes,
or equivalently, ${\rm Tr}[F(t,t')\rho(t')]$ being a backward
time invariable. The property is very analogous to that of the
Chapman-Kolmogorov equation in the classical diffusion
theory~\cite{Gardiner}. According to our previous experience
which constructing a more general backward time invariable may
lead into the classical fluctuation relations~\cite{LiuFJPA10},
it would be very interesting to explore whether the same idea
is still true here. Imitating Eq.~(5) in Ref.~\cite{LiuFPRE09},
we find there is very analogous backward time invariable in the
quantum case:
\begin{eqnarray}
{\rm Tr}[F\overline{\rho}(t)]={\rm Tr}[{\overline
F}(t,t')\overline{\rho}(t')],\label{BackwardInvariable}
\end{eqnarray}
if the new operator $\overline{F}(t,t')$ satisfies
\begin{eqnarray}
i\hbar\partial_{t'}\overline{F}(t,t')=-[\overline{F}(t,t'),
H(t')]-\overline{F}(t,t')(i\hbar\partial_{t'}\overline\rho(t')+[\overline\rho(t'),H(t')]){\overline{\rho}}^{-1}(t')\nonumber\\
+([\overline{F}(t,t'),A(t')]B(t')+\overline{F}(t,t')[B(t'),A(t')]){\overline{\rho}}^{-1}(t')
\label{PerturbedKsEq}
\end{eqnarray}
and its terminal condition at $t$ is assumed to be $F$, where
the operators $A(t')$, $B(t')$ and invertible density operator
$\overline{\rho}(t')$ are arbitrary. The proof of
Eq.~(\ref{BackwardInvariable}) is straightforward. The meaning
of the last two terms on the right hand side will appear when
we chooses $\overline{\rho}(t')$ to be the system's density
operator ${\rho}(t')$, i.e. the term
$i\hbar\partial_{t'}\overline{\rho}+[\overline{\rho},H]$
vanishing. The general Eq.~(\ref{PerturbedKsEq}) seem uncommon
in the quantum mechanics except a specific case:
\begin{eqnarray}
\label{quantumFKequation}
\left\{%
\begin{array}{ll}
i\hbar\partial_{t'}\overline{F}(t,t')=-[\overline{F}(t,t'),
H(t')]-\overline{F}(t,t')O(t'), &  \\
   \overline{F}(t,t'=t)=F,
   \end{array}
\right.
\end{eqnarray}
where $O(t')$ is an arbitrary operator. We may easily write
down its solution given by
\begin{eqnarray}
\label{quantumFKformula}
\overline{F}(t,t')&=&U(t')F^{\rm H}(t){\cal T}_{+}e^{{(i\hbar)}^{-1}\int_{t'}^t
d\tau U^{\dag}(\tau)O(\tau)U(\tau)}U^{\dag}(t')\\
&=&U(t')F^{\rm H}(t)Q(t,t')U^{\dag}(t'),
\end{eqnarray}
where ${\cal T}_{+}$ is the time-ordering operator. We simply
name Eq.~(\ref{quantumFKformula}) quantum FK formula because of
its highly formal similarity to the classical FK
formula~(\ref{FKformula}). However, we must remind the reader
that the whole time-ordering term, which we specially denote it
by a operator $Q(t,t')$ for convenience, only indicates that
the operator satisfies
\begin{eqnarray}
\label{TimeOrderingEq}
i\hbar\partial_{t'}Q(t,t')=-Q(t,t')[U^{\dag}(t')O(t')U(t')]
\end{eqnarray}
with a terminal condition $Q(t,t'$$=$$t)$$=$$1$. So far, we
mainly concentrate on a formal development; the physical
relevance of the quantum FK formula~(\ref{quantumFKformula})
and the backward time invariable~(\ref{BackwardInvariable}) is
not obvious. In the following we show that these results would
lead into the quantum Jarzynski and Bochkov-Kuzovlev equalities
if one chose specific $\overline\rho(t')$, $A(t')$, and $B(t')$.\\

{\it Quantum Jarzynski equality.}  We assume that the closed
quantum system is initially in thermal equilibrium with a
density operator $\rho_{\rm eq}(0)$=$e^{-\beta H(0)}/Z(0)$,
where the partition function $Z(0)$=${\rm Tr}[e^{-\beta
H(0)}]$=$e^{-\beta G(0)}$, $\beta$ is inverse temperature and
$G(0)$ is the initial free energy. At later times the system
evolutes under the time-dependent Hamiltonian $H(t)$. We choose
$A$=$B$=0, $\overline{\rho}(t')$ to be the instant equilibrium
state $\rho_{\rm eq}(t')$=$e^{-\beta H(t')}/Z(t')$ with the
instant partition function $Z(t')$=${\rm Tr}[e^{-\beta
H(t')}]$=$e^{-\beta G(t')}$. Equation~(\ref{PerturbedKsEq})
then becomes
\begin{eqnarray}
\label{JEequation}
i\hbar\partial_{t'}\overline{F}(t,t')=-[\overline{F}(t,t'),
H(t')]-i\hbar\overline{F}(t,t')\partial_{t'}\rho_{\rm
eq}(t'){\rho_{\rm eq}}^{-1}(t').
\end{eqnarray}
Obviously, the above equation follows the structure of
Eq.~(\ref{quantumFKequation}) and especially
\begin{eqnarray}
O(t')=i\hbar[\partial_{t'}e^{-\beta H(t')}e^{\beta H(t')}+\beta \partial_{t'}G(t')]
\end{eqnarray}
when we substitute the expression of $\rho_{\rm eq}(t')$ into
the ``source" term of Eq.~(\ref{JEequation}). Intriguingly, in
this case Eq.~(\ref{TimeOrderingEq}) indeed has a very simple
analytical solution
\begin{eqnarray}
\label{JESourcSolution}
Q(t,t')=[U^\dag(t) e^{-\beta H(t)}U(t)]U^\dag(t')e^{\beta H(t')}U(t')]e^{\beta[G(t)-G(t')]}.
\end{eqnarray}
Hence, on the basis of
Eqs.~(\ref{BackwardInvariable}),~(\ref{quantumFKformula}),
and~(\ref{JESourcSolution}) we obtain
\begin{eqnarray}
{\rm Tr}[F\rho_{\rm eq}(t)]&=&{\rm Tr}[U(t')F^{\rm
H}(t){\cal T}_{+}e^{ \int_{t'}^td\tau U^{\dag}(\tau)
\partial_\tau\rho_{\rm eq}(\tau) \rho_{\rm eq}^{-1}(\tau)U(\tau) }U^\dag(t')\rho_{\rm eq} (t')]\label{orgquantumJE} \\
&=&{\rm Tr}[F^{\rm
H}(t)e^{-\beta H^{\rm H}(t)}e^{\beta  H^{\rm H}(t')}U^\dag(t')\rho_{\rm eq}(t')U(t')]\hspace{0.1cm}e^{\beta [G(t)-G(t')]}.
\label{quantumJE}
\end{eqnarray}
If $F$=1 and $t'$$=$$0$ Eq.~(\ref{quantumJE}) is just the
quantum Jarzynski equality on the inclusive
work~\cite{TalknerPRE07}:
\begin{eqnarray}
\langle{e^{-\beta H^{\rm H}(t)}e^{\beta H(0)}}\rangle_{\rm eq}(0)=e^{-\beta \Delta G(t)},
\end{eqnarray}
where $\Delta G(t)$=$G(t)-G(0)$, and we have used $\langle$
$\rangle_{\rm eq}$(0) to denote an average over the initial
density operator $\rho_{\rm eq}(0)$. Additionally,
Eq.~(\ref{quantumJE}) at $t'$$=$$0$ is also a specific case of
the general functional relation given by Andrieux and Gaspard
earlier; see Eq.~(12)
therein~\cite{AndrieuxPrl08}.\\

{\it Bochkov-Kuzovlev equality.} Here we consider a special
realization of the time-dependent Hamiltonian~\cite{Kubo}: a
dynamic perturbation $H_1(t)$ ($t$ $\ge$ 0) is applied on a
system that is initially in thermal equilibrium with a
time-independent $H_{\rm o}$, that is, the total Hamiltonian at
later times is $H_{\rm p}(t)$$=$$H_{\rm o}$$+$$H_1(t)$. We have
used the subscripts o and p to indicate ``perturbed" and
``original", respectively. Obviously, the system's initial
density operator is $\rho_{\rm p}(0)$$=$$\rho_{\rm
o}$$=$$e^{-\beta H_{\rm o}}/Z_{\rm o}$, where the partition
function $Z_{\rm o}$$=$${\rm Tr}[e^{-\beta H_{\rm
o}}]$$=$$e^{-\beta G_{\rm o}}$. Choosing $H(t')$=$H_{\rm o}$,
$A(t')$$=$$-H_1(t')$,
$B(t')$$=$$\overline\rho(t')$$=$$\rho_{\rm o}$ in
Eq.~(\ref{PerturbedKsEq}), we obtain the following equation
\begin{eqnarray}
\label{B-KequationA}
i\hbar\partial_{t'}\widetilde{F}(t,t')&=&-[\widetilde{F}(t,t'),
H_{\rm o}]-([\widetilde{F}(t,t'),H_1(t')]-\widetilde{F}(t,t')[\rho_{\rm o},H_1(t')]\rho_{\rm o}^{-1}\nonumber\\
&=&-[\widetilde{F}(t,t'),
H_{\rm p}(t')]-\widetilde{F}(t,t')[\rho_{\rm o} ,H_{1}(t')]\rho_{\rm o}^{-1}.
\end{eqnarray}
We have used a new symbol $\widetilde{F}$ to distinguish it
from the previous $\overline{F}$ because they satisfy different
equations. We see that the above equation also follows the
structure of Eq.~(\ref{quantumFKequation}), and especially
\begin{eqnarray}
O(t')=[e^{-\beta H_{\rm o}},H_1(t')]e^{\beta H_{\rm o}}.
\end{eqnarray}
It would be interesting to check whether there is a simple
analytical solution to Eq.~(\ref{TimeOrderingEq}) under this
circumstance. We find that it indeed has:
\begin{eqnarray}
\label{BKSourcSolution}
Q(t,t')=[U^\dag_{\rm p}(t) e^{-\beta H_{\rm o}}U_{\rm
p}(t)][U^\dag_{\rm p}(t') e^{\beta H_{\rm o}}U_{\rm
p}(t')].
\end{eqnarray}
Using
Eqs.~(\ref{BackwardInvariable}),~(\ref{quantumFKformula}),
and~(\ref{BKSourcSolution}) we establish another equality given
by
\begin{eqnarray}
{\rm Tr}[F\rho_{\rm o}]&=&{\rm Tr}[U_{\rm p}(t')(F)_{\rm p}^{\rm
H}(t){\cal T}_{+}e^{(i\hbar)^{-1} \int_{t'}^td\tau U_{\rm p}^{\dag}(\tau)[\rho_{\rm o},H_1(\tau)]\rho_{\rm o}^{-1}U_{\rm p}(\tau)}
U_{\rm p}^\dag(t')\rho_{\rm o}]\label{orgquantumBK} \\
&=&{\rm Tr}[(F)_{\rm p}^{\rm
H}(t)e^{-\beta (H_{\rm o})_{\rm p}^{\rm H}(t)}e^{\beta  (H_{\rm o})_{\rm p}^{\rm H}(t')}U_{\rm p}^\dag(t')\rho_{\rm o}
U_{\rm p}(t')].
\label{quantumBK}
\end{eqnarray}
where $(F)^{\rm H}_{\rm p}(t)$=$U^{\dag}_{\rm p}(t)FU_{\rm
p}(t)$. If $F$=1 and $t'$$=$$0$ Eq.~(\ref{quantumBK}) is the
quantum Bochkov-Kuzovlev equality on the exclusive work
\begin{eqnarray} \label{B-Kequality}
\langle e^{-\beta {(H_{\rm o})}_{\rm p}^{\rm H}(t)}
e^{\beta H_{\rm o}}\rangle_{\rm o}=1
\end{eqnarray}
that was proposed very recently in Ref.~\cite{CampisiPTRS11},
where $\langle$ $\rangle_{\rm o}$ indicates an average over the
initial density operator $\rho_{\rm o}$.

On the other hand, a distinction between the original and
perturbed systems is not absolute. In physics we may regard
$H_{\rm p}(t)$ as the original quantum system while $H_{\rm o}$
is perturbed system if the dynamic perturbation of the latter
is thought to be $-H_{1}(t)$. Exchanging the subscripts p and o
and changing the sign before $H_1$ into minus in
Eq.~(\ref{B-KequationA}), we then obtain another equation given
by
\begin{eqnarray}
\label{Kuboequation}
i\hbar\partial_{t'}\widehat{F}(t,t')&=-[\widehat{F}(t,t'),
H_{\rm o}(t')]+\widehat{F}(t,t')[\rho_{\rm p} ,H_{1}(t')]\rho_{\rm p}^{-1}.
\end{eqnarray}
Of course, the argument can be as well strictly proved by
choosing $H(t')$$=$$H_{\rm p}(t')$, $A(t')$$=$$H_1(t')$, and
$B(t')$$=$$\overline\rho(t')$$=$$\rho_{\rm p}(t')$ in
Eq.~(\ref{PerturbedKsEq}). We are interested in what new
equalities like Eqs.~(\ref{orgquantumBK}) and (\ref{quantumBK})
will be yielded. Doing analogous derivations, we have the
following results:
\begin{eqnarray}
{\rm Tr} [F\rho_{\rm p}(t)]&=&{\rm Tr}[U_{\rm o}(t')F^{\rm I}(t){\cal
T}_{+}e^{- (i\hbar)^{-1}\int_{t'}^t d\tau U^{\dag}_{\rm o}(\tau)
[\rho_{\rm p}(\tau),H_1(\tau)]\rho^{-1}_{\rm p}(\tau) U_{\rm o}(\tau)}U^\dag_{\rm o}(t')\rho_{\rm p}(t')]
\label{orgquantumKuboEq}\\
&=&{\rm Tr}[F^{\rm
I}(t)(\rho_{\rm p})^{\rm I}(t)(\rho^{-1}_{\rm p})^{\rm I}(t')U_{\rm o}^\dag(t')\rho_{\rm p}(t')U_{\rm o}(t')],
\label{quantumKuboEq}
\end{eqnarray}
where $F^{\rm I}(t)$ and $\rho_{\rm p}^{\rm I}(t)$ are the
interaction pictures of the observable and density operator,
respectively, e.g. $F^{\rm I}(t)$$=$$U^{\dag}_{\rm o}(t)FU_{\rm
o}(t)$. Particularly, if $F$=1 and $t'$$=$$0$,
Eq.~(\ref{quantumKuboEq}) becomes
\begin{eqnarray} \label{Kuboequality}
\langle (\rho_{\rm p})^{\rm I}(t)
e^{\beta H_{\rm o}}\rangle_{\rm o}=e^{\beta G_{\rm o}}.
\end{eqnarray}
So far, we are not very clear what physics the equality
reveals, though its first order approximation should be related
to the famous fluctuation-dissipation theorems~\cite{Kubo}.\\

{\it Time reversal.} In the remaining part we want to give time
reversal explanations of
Eqs.~(\ref{JEequation}),~(\ref{B-KequationA}),
and~(\ref{Kuboequation}), which is essential to understand the
physical meaning of the backward time $t'$ and the origin of
these three equations. We use Eq.~(\ref{JEequation}) as an
illustration. Multiplying its both sides with $\rho_{\rm
eq}(t')$ and introducing a parameter $s$$=$$t$$-$$t'$
(0$<$$s$$<$$t$), we rearrange the equation into
\begin{eqnarray}
\label{initialproblem}
i\hbar\partial_{s}[\overline{F}(t,t-s)\rho_{\rm
eq}(t-s)]=[\overline{F}(t,t-s)\rho_{\rm eq}(t-s), H(t-s)].
\end{eqnarray}
Noting this is an initial condition rather than the terminal
condition problem. Equation~(\ref{initialproblem}) seems very
analogous to an evolution equation of a density operator, which
is indeed true if we multiply both sides of the equation by the
antiunitary time-reversal operator $\Theta$ and its conjugation
to obtain
\begin{eqnarray}
\label{reversedJEequatioon}
i\hbar\partial_{s}\overline{\rho}_{\rm R}(s)=[\overline{H}_{\rm R}(s),\overline{\rho}_{\rm R}(s)],
\end{eqnarray}
where the time-reversed density operator and time-reversed
Hamiltonian are
\begin{eqnarray}
\label{TimeReversedDensityH}
\overline{\rho}_{\rm R}(s)&=&\frac{1}{{\rm Tr}[F\rho_{\rm
eq}(t)]}\Theta\overline{F}(t,t-s)\rho_{\rm
eq}(t-s)\Theta^{\dag},\label{reverseddensity}\\
{\overline H}_{\rm R}(s)&=&\Theta H(t-s)\Theta^\dag,
\end{eqnarray}
respectively, and the coefficient is for the normalization of
$\overline{\rho}_{\rm R}(s)$.
Equation~(\ref{TimeReversedDensityH}) immediately explains the
backward time invariable in the Jarzynski equality case.
Additionally, one may also prove that the same equation is
equivalent to the key Lemma in Ref.~\cite{AndrieuxPrl08} that
was used to prove the functional relation; see the Appendix.
Doing very similar calculations, we find that
Eq.~(\ref{B-KequationA}) is equivalent to the evolution
equation of the time-reversed density operator
$\widetilde{\rho}_{\rm R}(s)$ with Hamiltonian
$\widetilde{H}_{\rm R}(s)$, which are given by
\begin{eqnarray} \widetilde{\rho}_{\rm R}(s)&=&\frac{1}{{\rm
Tr}[F\rho_{\rm
o}]}\Theta\widetilde{F}(t,t-s)\rho_{\rm o}\Theta^{\dag},\label{reverseddensity}\\
\widetilde{H}_{\rm R}(s)&=&\Theta H_{\rm p}(t-s)\Theta^\dag=H_{\rm
o}+\Theta H_1(t-s)\Theta^\dag\label{reversedH},
\end{eqnarray}
respectively, and Eq.~(\ref{Kuboequation}) is equivalent to
another evolution equation, the time-reversed density operator
and Hamiltonian of which are
\begin{eqnarray} \widehat{\rho}_{\rm R}(s)&=&\frac{1}{{\rm
Tr}[F\rho_{\rm
o}]}\Theta\widehat{F}(t,t-s)\rho_{\rm p}(t-s)\Theta^{\dag},\label{reverseddensity}\\
\widehat{H}_{\rm R}(s)&=&\Theta H_{\rm o}\Theta^\dag=H_{\rm
o},\label{reversedH}
\end{eqnarray}
respectively. Comparing the above three time reversal
explanations in the same dynamic perturbation problem, we see
that, though the original and time-reversed Hamiltonian for the
quantum Jarzynski equality is completely the same with that for
the quantum Bochkov-Kuzovlev equality, the time-reversed
density operators have very distinct initial conditions, i.e.,
$\overline{\rho}_{\rm R}(0)$$=$$\rho_{\rm eq}(t)$ and
$\widetilde{\rho}_{\rm R}(0)$$=$$\rho_{\rm o}$. Intriguingly,
for the equality~(\ref{Kuboequality}), the time-reversed
Hamiltonian is overlapped with the original one, both of which
are the unperturbed $H_{\rm o}$, while the initial conditions
of the time-reversed density operator and the original one are
different: they are $\rho_{\rm p}(t)$ and $\rho_{\rm o}$,
respectively.
\\

{\it Conclusions.} In this work, we have used a quantum
mechanical analogue of the classical FK formula to derive known
quantum nonequilibrium work relations in isolated quantum
systems. Compared with the previous methods in the literature,
our method shows highly similar in form to that for the
classical fluctuation relations which we developed
earlier~\cite{LiuFJPA10}. We think that it is insightful
because one may find the backward time invariable first and
then give its physical interpretation rather than vise versa.
Previous work has shown that direct defining nonequilibrium
physical quantities was very nontrivial in quantum
case~\cite{Campisi}. Extending our method into more complicated
quantum systems, e.g. the open quantum
systems would be a challenge in our following researches.\\

{\noindent 
This work was supported in part by the National Science
Foundation of China under Grant No. 11174025.}

\section{Appendix} Equation~(\ref{TimeReversedDensityH}) could be
further simplified. the time reversed density operator at later
time $s$ is connected with the initial condition by
\begin{eqnarray}
{\overline\rho}_{\rm R}(s)={U}_{\rm R}(s)\overline{\rho}_{\rm R}(0){U}_{\rm R}^{\dag}(s)
=\frac{1}{{\rm Tr}[F\rho_{\rm
eq}(t)]}{U}_{\rm R}(s)\Theta F\rho_{\rm eq}(t)\Theta^{\dag}{U}_{\rm R}^{\dag}(s),
\end{eqnarray}
where $U_{\rm R}(s)$ is the time-evolution operator for the
time-reversed Hamiltonian $\overline{H}_{\rm R}(s)$.
Substituting the above equation and the solution of
Eq.~(\ref{JEequation})
\begin{eqnarray}
\overline F(t,t')=U(t')F^{\rm H}(t)e^{-\beta H^{\rm H}(t)}e^{\beta H^{\rm
H}(t')}U^{\dag}(t')e^{\beta [G(t)-G(t')]}
\end{eqnarray}
into Eq.~(\ref{TimeReversedDensityH}) and doing a simple
calculation we obtain
\begin{eqnarray} {U}_{\rm
R}(s)=\Theta {U}(t-s){U}^{\dag}(t)\Theta^{\dag}.
\end{eqnarray}

\end{document}